# Transmembrane and Antimicrobial Peptides. Hydrophobicity, Amphiphilicity and Propensity to Aggregation

**M. Pirtskhalava**[*] (*m.pirtskhalava@lifescience.org.ge*), **B. Vishnepolsky** (*b.vishnepolsky@lifescience.org.ge*), **M. Grigolava** (*maia.grigolava@science.org.ge*)

I. Beritashvili Center of Experimental Biomedicine, Tbilisi 0160, Georgia

[*]*Corresponding author: m.pirtskhalava@lifescience.org.ge, Tel:: (995) 574 162397*

*Keywords:* antimicrobial peptides, transmembrane peptides, hydrophobicity, amphiphilicity, aggregation, comparative overview

**Development of the new antimicrobial agents against antibiotic resistance pathogens is the nowadays challenge. Antimicrobial peptides (AMP) occur as important defence agents in many organisms and offer a viable alternative to conventional antibiotics. Therefore they have become increasingly recognized in current research as templates for prospective antibiotic agents. The efficient designing of the new antimicrobials on the basis of antimicrobial peptides requires comprehensive knowledge on those general physical-chemical characteristics which allow to differ antimicrobial peptides from non-active against microbs ones.**
**According to supposed mechanisms of action, AMP interact with and physically disrupt the bacterial membranes. Consequently, hydrophobicity, amphiphilicity and intrinsic aggregation propensities are considered as such major characteristics of the peptide, which determine the results of peptide-membrane interactions. For some kind of peptides such characteristics as hydrophobicity, amphiphilicity and aggregation bias determines their ability to compose transmembrane domain of the membrane protein, whilst for others the same properties are respond for their antimicrpobial activity, i.e. give them ability of membrane permeability and its damage. In this review we analyze the data about hydrophobicity, amphiphilicity and intrinsic aggregation propensities available in literature in order to compare antimicrobial and transmembrane peptides and show what's the common and what's the difference in this respect between them.**

## Introduction

It is generally recognized that overall hydrophobicity is the main driving force for the integration of trans-membrane (TM) segments of the protein into the lipid bilayer. Transmembrane proteins have a signal sequence that directs the ribosomally-syntesized polypeptide into the large transmembrane translocon for a final folding and translocation in order to organize native membrane embedded structure [1,2]. That is, transmembrane fragments of membrane protein chain are "aggregated" (folded) in order to form final, stable structure of transmembrane domain without dramatic perturbation of lipid bilayer. Most of the antimicrobial peptides are also synthesized as a precursor's chain on the ribosome and then matured and secreted onto extracellular environment. Although ribosomally-synthesized antimicrobial peptides are unordered in extracellular environment and don't show any aggregation propensity, it's considered that in the lipid environment they are structured and propensity to aggregation arises. The result of antimicrobial peptide aggregation is the pore-formation or even membrane structure disruption. That is, transmembrane and antimicrobial peptides resemble each other in many aspects. But, at the same time, the results of their interaction with membrane are different. The functional differences between these two membrane interacting peptides should be expressed in their amino acid sequences.

Lipid bilayer of cell membrane is highly heterogeneous in the normal to the membrane plain direction. The membrane resembles a sandwich with internal hydrophobic part, enclosed by two interfacial water enriched layers (in contact with intracellular and extracellular environment) [3]. Consequently, the membrane is amphiphilic in nature and compounds having the obligation to associate with, embed into, or permeate the membrane should be hydrophobic or amphiphilic. Antimicrobial peptides (AMP) are relatively short, mainly positively charged and amphiphilic and so have all features necessary to interact with bacterial and fungal cell membrane. In general amphiphilic nature is characteristic of many membrane bound peptides and putative transmembrane helices (TMH) of membrane proteins [4, 5]. At the same time, a hydrophobicty and amphiphilicity are major players directing aggregation as well as folding of polypeptides either in polar or membrane environment. Hydrophobicty and amphiphilicity of peptide are defined by composition and distribution along the chain of the residues of particular physical-chemical nature. So, a knowledge of sequence characteristics such as hydrophobicity, amphiphilicity and intrinsic aggregation propensities are needed for revealing a function of query peptide  This is the reason why the last characteristics estimations are so widely appeared in the literature. Recent work suggests that antimicrobial activity is not dependent on specific amino acid sequences or on specific peptide structures (6). Instead, activity depends more on the amino acid composition of a peptide and on its physical-chemical properties. In this work we try to overview occurred in literature data about amino acid composition and residues distribution which defines hydrophobic, amphiphilic and intrinsic aggregation propensities of transmembrane and antimicrobial peptides and answer the question **what 's the common and what's the difference in this respect between these two different classes of membrane-interacting peptides.**

# Hydrophobicity

A hydrophobicity of a chemical compound is related to its transfer free energy from a polar medium (phase) to non-polar medium (phase). Hydrophobicity of a peptide is usually evaluated as a sum of particular amino acid transfer free energies. A hydrophobicity could allow to understand how polypeptide chain forms and stabilizes their 3D structure in polar or non-polar environment or how it interact with membranes during key biological events. Ranked collection of the water-to-bilayer transfer energies of the twenty natural amino acid side chains is called a " hydrophobicity scale".

**Hydrophobicity Scales.** The first scale was proposed by Nozaki and Tanford in 1971 and today more than 100 different scales are offered [7-29]. Scale can be experimental or knowledge-based (calculated). There are sensible differences between scales. These differences can be explained by differences in the type of method used and conditions in which they realized. For example: the two incapable of being mixed phases can be either a system composed of hydrophilic and hydrophobic isotropic solvents (water-octanol, water-cyclohexane, etc) or a system composed of water and anisotropic structures such as micelle, liposome, membrane, etc.; at the same time, biophysical properties of membranes and their hydration vary; the different carriages by which amino acid transitions between phases take place can be used (for instance: peptides, proteins, translocon machinery, etc.); the methods used to estimate transfer free energy are also different: HPLC-based, partitioning-based, accessible surface area-based, site-directed mutagenesis based, physical property-based, etc.; the conditions (pH, ionic strength, etc.) in which transition takes place also vary. Therefore, it is logical to think that it's impossible to find a single transfer free energy (hydrophobicity) scale that is optimal under all circumstances.

Recently the comparison of hydrophobicity scales based on experiments with increasing biological complexity has been done by J.L.MacCallum & D.P.Tieleman [30]. The simplest scale considered by authors was Radzicka–Wolfenden small molecule partitioning scale [31] based on partitioning of small molecules between bulk solvents. The most complex scale was Hessa and coworkers [26] based on *in vitro* experiments on the insertion of a transmembrane helix by the Sec translocon. As the scale of middle complexity were considered the Moon–Fleming OmpLA folding/refolding scale [32], the Wimley–White pentapeptide-based hydrophobicity scales [21-24] and the MacCallum et al. molecular dynamics potential of mean force scale [33]. The absolute magnitudes of the Wimley–Hessa–Moon scales and the MacCallum–Radzicka scales differ by significant amount, but at the same time different scales are highly correlated. For example Hessa et al. scale is based on a biological system and therefore it has been suggested that it does not measure a true thermodynamic equilibrium. However, as showed [30] the Hessa scale is highly correlated with the other scales. Authors have concluded: "The differences cannot arise from simple constant shifts in free energy, as would be expected if the differences were due to overall differences in the hydrophobicity of the phases used in each experiment. Rather, the scales differ by a multiplicative factor: if the polar residues have a higher free energy in one scale relative to another, then the non-polar residues will have a more favorable free energy of transfer so that the ratio of free energies is similar".

Thus, taking into account diversity of scales the evaluation of the peptide behaving in water, membrane, or membrane-water interface environment is problematic. As a rule, polypeptide –membrane interactions are characterized by free energy of transfer of polypeptide from water environment into membrane. The hydrophobic part of this free energy is usually evaluated as a sum of particular, constituted peptide amino acids transfer free energies. The satisfactory approximation of the free energy of transfer of polypeptide depends on the right selection of the hydrophobicity scale. The selection should be done carefully and the non-additivity effect demonstrated for only arginines [34] ought to be taken into account.

**Hydrophobicity and transmembrane segments of protein**
Segrest & Feldman [35] and Rose [36] noted that numerical hydrophobicities might be effective in detecting hydrophobic segments of transmembrane protein. Kyte & Doolittle [37] found that averaging hydrophobicity over segments of 19 residues is most effective in distinguishing membrane-spanning segments from globular controls. Eisenberg et al [38] had concluded that it may be necessary to consider cooperativity to be able to identify membrane-spanning sequences. The authors used the normalized consensus hydrophobicity scale [12] and averaged over a segment of 21 residues that moves through each of the sequences.

**Knowledge-based potentials.** The neutron diffraction studies have shown that water-lipid interface to be continuous rather than discrete [39]. Knowledge-based hydrophobicity scales are generally based on consideration that the bilayer consists of distinct sectors representing the hydrophobic hydrocarbon core, a polar headgroup region, and the extramembrane aqueous region. The difficulties risen by such consideration and connected with the impossibility to estimate exactly the boundaries of bilayer sectors would be the cause of uncertainty in transfer energy evaluations. In order to avoid this difficulty and account heterogeneity of the lipid-water interface, Martin B. Ulmschneider et al. [40] have developed the potential that describes the free energy profile for each residue type as a continuous function of its depth of insertion within a membrane. The study involved 46 α-helical inner membrane proteins (TMH) containing 440 non-redundant TM helices. Distributions of each amino acid in the trans-membrane domain were calculated as a function of the membrane normal and potentials of mean force along the membrane normal were derived for each amino acid by fitting Gaussian functions to the residue distributions. The individual potentials agree well with experimental and theoretical considerations. For example the potential describes the inside/outside orientation of the membrane proteins correctly.

Recently, related, a depth-dependent insertion potential, Ezα, for TMH was reported by Senes at al. and coworker [41]. They analyzed helical inner membrane proteins and determined depth-dependent propensity profiles for amino acid side-chains. The N-terminal ends were not differentiated from the C-terminal ends of the helices and the distance of the residues $C^\beta$ ($C^\alpha$ for Gly) from the bilayer center was measured. With the exception of Trp and Tyr, propensity dependence on depth was fitted by sigmoidal function. For Trp and Tyr, whose propensities reach a maximum in the headgroup region and decrease in the membrane center, a Gaussian distribution was used for propensity distribution. The results of water–membrane [12, 24] and water–interface [21] transfer energy scales are correlated perfectly with the potentials. Translocon-mediated apparent free energies of insertion [26] are also in good agreement with predictions done by depth-dependent potential.
Although the methodologies used by Ulmschneider et al. [40] and Senes et al. [41] are different, the two potentials are in good agreement overall. However there are a few differences. The major difference between the two methods is His. According to Senes et al [54], this amino acid follows a sigmoidal distribution that favors partition into water, which is very similar to all other polar residues. In Ulmschneider's potential, His behaves like Tyr and Trp, with Gaussian curves that have minima in the interfacial region. This significant disagreement might depend on differences in the databases used.

Hsieh et al. [42] focused on the energetics of insertion of outer membrane β-barrel proteins (TMBs) into the lipid bilayer and developed a depth-dependent potential Ezβ for TMBs, which reflected the unique lipid environment, folding pathway and secondary structural context of this class of proteins. Ezβ was calculated using the same protocol as Ezα [41], based on thirty-five high-resolution TMB crystal structures of low sequence homology. A comparison of Ezα [41] and Ezβ [42] shows that most amino acids exhibited similar distributions. Values of the parameter for Ezb strongly correlated with those of Ezα (r = 0.78) and with an experimentally derived hydrophobicity scale [12] (r = 0.68), indicating that general physico-chemical behavior was conserved across both classes of membrane proteins [42]. That is, polar residues preferred the outside of the membrane, while hydrophobic residues had the reverse preference and aromatic residues were predominantly situated in the headgroup region.

Although values of the parameter for Ezα and Ezb correlate with an experimentally derived hydrophobicity scale they are carrier of useful additional information concerning the peptide's depth of insertion into the membrane and peptide orientation in relation to membrane surface. Depth and orientation in combination with charge and aggregation propensity are the major determinants of understanding of peptide action mechanism. So peptide's insertion depth and orientation can become principal parameters of designing when more and more transmembrane domain's 3D structures get into the PDB and more refined values of Ezα and Ezβ parameters are available.

# Amphiphilicity

Amphiphilic (amphipatic) compound is the one which consists of two, lipohpilic and hydrophilic parts and these parts are segregated from each other in the space. Therefore amphiphilic molecules prefer the interface between polar and non-polar environments.

The phospholipids have an amphiphilic character. The amphiphilic nature of these molecules defines the way in which they form membranes by arranging into bilayers and defining a non-polar region between two polar ones. That is, membrane is also of an amphiphilic structure.

Certain amino acids possess "innate" amphiphilic properties. On the other hand, polypeptide chain fragments can get an amphiphilic structure by means of certain distribution of polar and non-polar amino acids along the chain. Amphiphilicity is an essential factor directing protein folding and polypeptide-membrane association process. The amphiphilicity is a characteristic of many membrane bound peptides and putative transmembrane helices of membrane proteins [43-45,65]

There are some quantitative characteristics describing a level of amphiphilicity of particular amino acids or polypeptide fragments.

## Amphiphilicity scale

Some amino acids side chains are composed of polar group and large hydrophobic stem beyond γ–carbon atom and so originally are amphipatic in nature. Amino acids of such kind, as a rule are gathered in the membrane–water interface and act as a crucial actor in polypeptide-membrane association. Mitaku S et al. [46] in order to improve their membrane protein prediction algorithm have developed a novel index that represents the amphiphilicity of each amino acid side chain. Each Side chain is indexed according to the transfer energy, which was directly proportional of accessible surface area of the hydrophobic groups with the coefficient of proportionality 40 dyn $cm^{-1}$. Consequently transfer energy for aromatic and aliphatic stems were calculated and positive values were obtained for seven amino acids Lys, Arg, Hys, Glu, Gln, Trp, Tyr. Amphiphilicity index values of polar Asp, Asn, Ser and Tre amino acids are zero because their polar groups are connected to the main chain through β-carbon only. Since the hydrophobic amino acids have not polar groups their values of indexes are also zero.

**Amphiphilic helix. Hydrophobic Moment**

It was found out that most of the secondary structures (α-helix, β–strucutre) of proteins are amphiphilic. i.e. one surface of secondary structure is hydrophilic and other is hydrophobic. The helix amphiphilicity was represented by Schiffer & Edmundson [47] as two-dimensional "helical wheel" diagrams, a projection down the helix axis showing the relative orientations of residues. A quantitative measure of the amphiphilicity perpendicular to the axis of the α– helix or β–structure segments, was proposed by Eisenberg et al [48] and called the hydrophobic moment. It can be calculated for an amino acid sequence of $N$ residues on the basis of transfer free energy of i-th residue $h_i$ by the equation

$$M = 1/N([\sum_i (h_i \cos(ia))]^2 + [\sum_i (h_i \sin(ia))]^2)^{1/2}$$

where the sums are over all residues of the peptide, and $a$ is the angle in radians, which corresponds to periodicity of structure. For the α-helix, $a \approx 100$ and for β–strand $a \approx 180$. If hydrophobic and hydrophilic residues are evenly distributed among the helix hydrophobic moment is small, whilst if the most of the hydrophobic residues are placed on one side of the helix and most of the hydrophilic residues on the other the moment becomes large. Thus, the hydrophobic moment measures the extent of amphiphilicity of a helix.

**Hydrophobic moment plot**

To quantify the amphiphilicity of protein secondary structures, Eisenberg and co-workers [48] introduced the hydrophobic moment, $M(a)$ and developed hydrophobic moment plot methodology, which provides a graphical technique for the general classification of protein α-helices [49]. The hydrophobic properties of a helix can be represented as a point on a hydrophobic moment plot, on which the vertical axis is the hydrophobic moment per residue, and the horizontal axis is the hydrophobicity per residue. The different classes of proteins have the tendency to plot in different regions of the hydrophobic moment plot. For examples, globular proteins generally plot in the certain region, at intermediate values of hydrophobicity and hydrophobic moment, whilst surface-seeking proteins plot in the region with a high hydrophobic moment [49].

**Linear moment**

The results of analysis of transmembrane helix sequences don't show uniform distribution of amphiphilic and hydrophobic residues along the polypeptide chain line. For example Mitaku et al. [46] have shown that there is general tendency for transmembrane helix to have a significant peak in hidrophobicity in the central region whilst peaks in amphiphilicity value occur very near the end of helices. However they noted, that although profiles of strongly polar (R,K,E,H,Q) and weakly polar residues (W,Y) show high value of amphiphilicity at the end points of transmembrane helices, the peak in values of strongly polar residues is located just outside of the helix, whereas the peak in values of weakly polar residues is apparently located inside the helix.

Ulmschneider and coworkers [40] also noted that although charge residues account for less than 6% of all the residues in the TM domain, they make up one fifth (19%) of residues in the interfacial regions. Membrane proteins generally have an asymmetric charge distribution along the membrane normal. This provides for the correct orientation of the protein in the membrane as well as preventing the loss of the protein to the extracellular space. Indeed Ulmschneider and coworker show that there is a bulk of charge on the exposed side of the membrane protein. The net charge along the membrane normal was calculated by assuming all ionizable residues (Arg, Asp, Glu, Lys) with a surface accessibility greater than 10% to be charged, while all others were taken to be neutral. The averaged net total charge per protein on the "inside" (i.e., cytoplasm, matrix or stroma) was found to be +3.8 ± 0.2 $e$, compared to -4.5 ± 0.2 $e$ on the "outside", giving strong support to the "positive-inside rule" [40]. Thus, for every three positive residues on the intracellular side there are four negative residues on the extracellular side [40].

Therefore it is reasonable to suppose the existence of linear separation of hydrophobic and hydrophilic residues along the chain in transmembrane peptides, i.e. we can consider linear aphiphilicity of the peptide, for which the helical hydrophobic moment [45] is not useful as the measure of aphiphilicity and consequently new quantitative measure is needed for such case. Vishnepolsky ans Pirtskhalava [50] have developed the new quantitative characteristic for linear amphipaticity, the linear hydrophobic moment which is calculated as

$$M = D(\sum h_k^+ - \sum h_k^-),$$

where

$$D = |\sum h_k^+ \cdot k / \sum h_k^+ - \sum h_k^- \cdot k / \sum h_k^-|$$

D is a distance between the centers of hydrophobic and hydrophilic part of the peptide of length N; k=1,N, $h_k^+$ and $h_k^-$ - transfer energy of the k-th residue from water to the hydrophobic environment under the conditions that $h_k^+ > 0$ corresponds to hydrophobic residue and $h_k^- < 0$ corresponds to hydrophilic residue.

# Aggregation

The same Physical forces responsible either for forming secondary structure elements or stabilizing aggregates of two or more helices or beta strands. Folding of the polypeptide chain (i.e. aggregation their fragments) as a rule is a process of formation and interaction of secondary structure elements. This is a reason that the mechanisms of alpha helix or beta structure segments formation / interactions and the factors of their aggregates stabilization are well studied.

It has been suggested that in water environment hydrophobic interactions between non-polar side-chains probably contribute to the dominant stabilizing interactions for secondary structure formation and folds (their aggregates) stabilization. The mechanisms of the alpha aggregation (alpha helices interaction) and the beta aggregation (beta strands interaction) in water environment became the object of many investigations. The aim of this section is to overview the outcomes of these investigations.

**Alpha aggregation in Water (Polar) Environment**

As a model of alpha aggregation in water environment coiled coil structures are considered. Coiled coils are α-helical protein structural motifs responsible for multimerization. Left-handed coiled coil sequence show a characteristic seven-residue repeat ( *a b c d e f g* )$_n$, where *a* and *d* are typically nonpolar residues found at the interface of the helices and drives helices to associate, whereas *e* and *g* are solvent exposed polar residues. Thus, combination of three- and four-residue intervals for polar and non-polar amino acids is a prerequisite for coiled coil packing.

Despite this shared hydrophobic/hydrophilic pattern, coiled-coil sequences adopt different conformation (dimeric, trimeric and tetrameric aggregates) and type of conformation depends on buried residues (in *a* and *d* positions) type. The shapes of buried amino acids in coiled coil are essential determinants of global fold. Results of several investigations [51-54] show, that the type of coiled-coil oligomerization is determined by the distribution of β-branched residues at *a* and *d* positions. The occurrence of β-branched residues at *d* positions disfavor dimers, while β-branched residues at *a* positions should disfavor tetramers. The presence of β-branched residues at both *a* and *d* positions facilitates trimer formation.

Helices pairing specificity is greatly influenced by the nature of the electrostatic *e* and *g* residues. The charge pattern on the outer contacting edges of a coiled coil dictates its preference for homo- or heterotypic pairing, and whether the orientation of the coiled coil is to be parallel or antiparallel [55,56].

Membrane occurrence can influence on the coiled coil formation in water. For example some peptides which are not α – helical and not be able to form a stable α -helical coiled-coil structure in solution, but have possibilities to be associated (not inserted) with membrane (positively charged residue in *f* position of heptad gives possibility to interact with anionic lipid vesicles (POPG)) form α-helices and consequently coiled coil at interaction with membranes [57].

In globular protein, interacting helices usually have not identical sequences and so interfacial composition of the interacting helices may be slight different. Kleiger and coworker [58] have revealed the GXXXG or AXXXA motif can promote the association of helices.

**Beta aggregation in water (Polar) environment**

The physical model of strand interactions, i.e. β–sheet formation, supposes backbone strong H-bond interactions and non-H-bonded side-chain interactions. Cross-strand interactions, i.e the closest contacts between the side chains do noticeable contribution into a beta sheet stabilization. Large aromatic residues (Tyr, Phe and Trp) and β-branched amino acids (Val, Ile) are favored to be found in β strands in the middle of β sheets. Glycine is an intrinsically destabilizing residue in β sheets. In natural proteins, however, this destabilization can be 'rescued' by specific cross-strand pairing with aromatic residues [59]. By using the binary code strategy Kamtekar et al constructed several libraries of *de novo* proteins [60]. Libraries based on the α -helical binary pattern, that is - combination of three- and four-residue intervals for polar and non-polar residues, indeed yield α-helical proteins (aggregates of the α-helical fragments); and libraries based on the β-sheet pattern, that is alternating patterns of polar and nonpolar residues indeed yield β-sheet proteins [60,61]. However, the properties of proteins from the two types of libraries differ dramatically. In contrast to α-helical proteins, the β-sheet proteins assemble intermolecularly into large oligomeric structures resembling amyloid fibrils. In general, regular β-sheet edges are dangerous, because they are already in the right conformation to interact with any other nearest β-strand. A structurally oriented analysis of how nature avoids β-strand aggregation was done by Richardson and Richardson [62] where they concluded, that many edge-protection strategies are used by natural proteins. One strategy used by nature is the avoidance of alternating patterns [63]. But the most useful strategy is the appearance of a single charged side chain near the middle of the hydrophobic side of the edge β-strand. Wang and Hecht [64] showed that incorporation of a lysine into the non-polar face of a β-strand of the de novo designed proteins can indeed prevent aggregation and favor monomeric β-sheet proteins.

## Transmembrane peptides

In membranes, the rules governing the energetics of secondary structure and fold formation are likely to be very different. For instance, hydrogen bonds become probably more important for driving secondary structure and their aggregates formation. The hydrogen-bond effect helps explain the easy formation of peptide secondary structure in membranes [65]. Wimley & White have shown that the free energy cost of moving a non-hydrogen-bonded peptide bond from water into the interface of a membrane is unfavorable by 1.2 kcal mol$^{-1}$ [21]. The simple fact that all trans-membrane domains of membrane proteins are either α-helices or β-barrels indicates that peptide bonds involved in hydrogen-bonded secondary structure have a much lower free energy than unrealized peptide bonds.

### A hydrophobicity and Amphiphilicity

As a rule helices from transmembrane proteins has a large values of hydrophobicity and low value of hydrophobic moment and so are situated at the certain, different from globular and surface-seeking proteins area of the hydrophobic moment plot [49]. But there are distinctions even among membrane-spanning helices. The single-chain membrane anchors plot in the region of highest hydrophobicity and lowest hydrophobic moment, whilst channel-forming membrane protein - at somewhat higher values of hydrophobic moment and lower values of hydrophobicity [66]. As the transmembrane helices have particular hydrophobic/amphiphilic features, it was suggested that the hydrophobic moment, as well as the hydrophobicity of peptides can yield some information about their propensities to association with or embedding into the membrane.

### Alpha aggregation

In membrane environment non-helical structure is unfavorable, but helix formation and their aggregation mechanism differ from that described for soluble proteins. Generally, environment is a key regulatory factor in determining the conformation adopted by a given protein sequence. The helical propensity of individual amino acids may be altered in response to the change in environment from water to the membrane. For instance the bulky residues, such as Ile and Val, which are ordinarily described as helix destabilizing, ranks as the best "helix-promoter" in membrane and it was found that they may be important for membrane protein assembly and folding [67]. Another examples are Gly and Pro, which destabilize the α-helices in globular proteins and polypeptides, but display a considerable tendency to form α-helices in membrane environments [67]. Consequently differences should be occurred in amino acid composition and sequence motives of "average" soluble and membrane proteins.

Generally, both membrane and water-soluble proteins commonly fold into bundles of α-helices. However, the composition and distribution of the amino acids in these proteins are very different. It's clear, that in soluble proteins polar and charged residues are on the water-accessible surface, whereas in membrane proteins hydrophobic residues cover the lipid-exposed surface. At the same time to understand how the helix interacts and aggregates the knowledge of the nature and distribution of amino acids in the interiors are needed. Compositional biases of the interiors of the water-soluble α-helical bundles we have reviewed in previous section. Here we'll try to concern an arrangement of interior of transmembrane α-helical proteins (domains).
It's well known, that polar-polar interactions are involved in interhelical contacts in soluble proteins as well as in membrane ones. However, the patterns of polar-polar contacts are different. In soluble proteins, only salt bridge, i.e. ionizable residue pairs have a high propensity to interacting helices, whereas in membrane proteins in addition to salt bridges, there are residue pairs between ionizable and polar residues. For instance polar residues (Asn, Asp, Glu or Gln) introduced in the TM helical segment can provide sufficiently strong driving force for their self-association in membrane environment [68,69]. Polar and ionizable residues form extensive H-bond connections between TM helices. Approximately all TM helices form one or more interhelical H-bonds. The helices contacting with H-bonds are packed tighter [70]. The small residues (Ala, Gly, Ser, Thr) provide high average packing values for helix interfaces of transmembrane domains. Generally, TM segments of membrane protein are characterized by high frequency of glycine (9%) [40]. It has been reported that glycines occur frequently at helix–helix interfaces and crossing points [71] and so may facilitate closer packing of TM helices [72,68]. Gly-mediated packing has been explored via a series of experiments on glycophorin A dimers, [73,74] which firmly established the essential role of Gly in the dimerization motif.
Senes et al [75] have performed an accurate analysis of the frequency of occurrence of all pairs and triplets of amino acids in a large non-homologous set of TM sequences, compared with their theoretical expectancies. Their results show: all pairs formed by two small residues (Gly, Ala and Ser) at register 4 exhibit bias to high occurrence in TM . The β-branched residues Ile and Val correlate very strongly also at register 4 (II4, IV4, VI4, VV4). Combinations of a small and a large residues at register i, i+4 are strongly disfavored. Most of the positively correlating pairs occur at separations i, i+1 and i+4, i.e. on the same face in α-helical conformation. So, many of the amino acid correlations that were found in Senes et al [75] analysis are readily interpretable in terms of helix-helix interaction patterns. Authors emphasize, that the relationships between pairs of larger and smaller residues suggested that the correlations were not limited to the pairs. Triplets containing the pair GG4 also show the strongest positive correlation and mostly in conjunction with Ile, Val or Leu at registers ±1 and ±2 with respect to the Gly. The "GG4 + β-branched" motif and its variations can be found in many available X-ray structures of helical transmembrane proteins [75].

Later Kim S et al [76] have described another TM sequence motif, the glycine zipper. The most significant glycine zipper sequence patterns are (G,A,S)XXXGXXXG and GXXXGXXX(G,S,T). These patterns contain a GXXXG motif, which is shown to be important in TM helix homodimers [69, 73-75]. The glycine zipper packing mode is distinct from the GXXXG dimerization motif that involves direct packing between the Gly faces [77]. In glycine zipper packing, the glycine zipper packs against a different face of the associated helix.

Finally we can say, that the driving forces for helices association in membrane environment are most likely van der Waals packing and polar interactions, such as hydrogen bonds, the latter being especially strong in the low polarity of the membrane interior. Small residues facilitate to additional $C^{\alpha}$-H…O hydrogen bonds formation [78].

**Beta aggregation**

The non-bonded polar amide and carbonyl groups are unfavorable in membrane environment. Consequently, the geometry of the β-strands and the necessity to form hydrogen bonds between polar amide and carbonyl groups of the polypeptide chain within the hydrophobic core of the membrane excludes that individual β-strands can exist in a lipid bilayer. This is a reason, that all known integral membrane proteins with transmembrane β-strands form barrel structures. Thus, hydrogen bonding between backbone N and O atoms is a main driving force of β-strand interaction (aggregation) in membrane environment, but the two other types of interactions contribute in strand aggregates stability also. This interactions are: "non-H-bonded interaction", that is between a cross-strand pair of residues that interact but do not share backbone H-bonds; and "weak H-bond", that is between a cross-strand pair of residues that share weak $C_\alpha$ …O H-bonds [79]. Analysis of Jackups and Liang reveal some preferred interaction motives [79]. For instance Gly-Aromatic and Aliphatic-Aromatic cross-strand pairs have much higher preferences for strong H-bonds. Aromatic-Aromatic and Gly-Aliphatic cross-strand pairs have higher preferences for non-H-bonded interactions. Polar-Aliphatic cross-strand pairs (strong H-bonds) and Polar-Aromatic and Aliphatic-Aliphatic cross-strand pairs (non-H-bonded interactions) behave differently.

The comparing TM and soluble proteins shows, only two motifs: G-V non-bonded pairs and G-P weak H-bonds appear in both [79]. In soluble β-sheets, polar-polar and hydrophobic-hydrophobic pairs have high propensities for strong H-bond and non-H-bonded pairings, while polar-hydrophobic pairs have low propensities. In the membrane proteins weak H bonds show an opposing trend [79]. The strongest motifs in TM β-barrels, W-Y non-H-bonded pairs and G-Y strong H-bonds, are not statistically significant in soluble sheets. The only favorable interaction motif found in both TM α-helices and TM β-barrels is G-F, a strong H-bond motif. Otherwise, there are very few similarities between the two protein families [79].

It's very interesting the role of the membrane in amyloid peptide (Aβ) aggregation. Abedini and Raleigh [80] speculated that Aβ exists in a β-sheet or random coil configuration when soluble extracellularly or in the cytoplasm, but converts to an α-helical structure upon membrane association. The α-helix formation occurring upon membrane association is directed by the glycine zipper region of Aβ. It has been proposed that the formation of α-helical structure by membrane-associated Aβ drives the formation of oligomers and therefore a local increase in Aβ concentration, and this localized Aβ concentration catalyzes β-sheet formation and amyloid formation [80]. α-helical intermediates have been observed preceding amyloid formation [81]. According to Kirkitadze et al [81], glycine zipper-driven oligomerization at membranes is required to reach the critical local Aβ concentration to initiate amyloid formation. It should be noted, that amyloid peptide holds antimicrobial activity [82] and so peculiarities of dependence of the Aβ structure and aggregation propensity on environmental conditions is useful for understanding the mechanisms of action of antimicrobial peptides.

**Antimicrobial peptides**

In the vast majority of cases, antimicrobial peptides (AMPs) are cationic (CAP) and kill microbes via mechanisms that predominantly involve interactions between the peptide's positively charged residues and anionic components of target cell membranes. These interactions can then lead to a range of effects, including membrane permeabilization, depolarization, leakage or lysis, resulting in cell death [83,84]. But there are also Anionic Antimicrobial peptides (AAP) which have been established as an important part of the innate immune systems of vertebrates, invertebrates and plants [85]. Membrane interaction appears key to the antimicrobial function of AAPs and so, these peptides generally adopt amphiphilic structures. AAPs 3D arrangements vary from the α-helical peptides of some amphibians to the cyclic cystine knot structures observed in some plant proteins. Some AAPs appear to use metal ions to form cationic salt bridges with negatively charged components of microbial membranes, but in many cases, the mechanisms underlying the antimicrobial action of these peptides are unclear or have not been elucidated [85]. According to Harris et al [85] AAPs may be induced or expressed constitutively and in some cases, antimicrobial activity appears to be a secondary role for these peptides with other biological activities constituting their primary role. Studies on scorpion toxins have suggested that AAPs may be relics from the early evolution of antimicrobial peptides and that in the course of time the enhancement of toxicity to microbes has become associated with increases in the overall positive charge of antibiotics and toxins [86]. Other authors have suggested that AAPs arose to complement CAPs, providing a response to microbes that had developed resistance to these latter peptides [87].

In any case, CAPs either structurally or functionally are well and widely studied in contrast to AAPs . So, below we present an overview of CAPs only.

Generally, water soluble and membrane active CAPs are relatively short, mainly positively charged and amphiphilic, and can protect the host organism against bacterial and fungal attacks by destroying the barrier function of the invading microbe's membrane. CAPs show substantial sequence diversity and consequently different behavior in water and membrane environment. From a biophysical viewpoint, the efficacy of the CAPs should depend on net charge, overall hydrophobicity, peptide chain length, and the degree of ordered structure both in aqueous and membrane environments.

**Structure and Aggregation**

Because CAP reaches target membranes through the aqueous phase, their properties in aqueous solution are important for their effects on membranes. Rina Feder and coworkers tried to understand how CAP organization in aqueous solution might affect the antimicrobial activity and conclude that CAP potency correlated well with aggregation properties. They have shown that aggregation can have dramatic consequences on antibacterial activity of the Dermaseptin-derived peptides. More potent against bacteria peptides were clearly less aggregated [88].

Recent artificial neural-network prediction models of antimicrobial peptides have found that peptide aggregation in solution indeed contributes to a low antimicrobial activity [89]. Interestingly, the addition of cationic residues to peptides has been shown to inhibit aggregation in solution while improving the antimicrobial potency at the same time [90].

Propensity to aggregation is determined by sequence and consequently structural features of CAP. Linear CAP, because of its short length, combination of non-high mean hydrophobicity with relatively high net charge shows the disordered structure in aqueous solution and prevention of the aggregation (high net charge and the resulting electrostatic repulsion between peptides limits aggregation). Cycled main chain or intrachain covalent bonds represent prerequisite for certain structural stability of CAP in aqueous solution and for possibility of aggregation.

**Gramicidin S (GS)** is a cyclodecapeptide antibiotic, constructed as two identical pentapeptides joined head to tail. The simple C2 - symmetric cyclic GS has significantly limited conformational flexibility, and its experimentally observed antiparallel β-sheet conformation with two Type II' β-turns is rigidly retained in solution [91] and crystal form [92] as well as in DPPC bilayers [93]. In the crystals grown in the presence of trifluoroacetic, the gramicidin S molecules line up into double-stranded helical channels [94]. Khalfa and Tarek have shown that arginine-rich cyclic peptide assembles into nanotubes and throws out the lipid bilayer [95].

**The protegrins (PG)** are a family of arginine- and cysteine-rich cationic peptides found in porcine leukocytes that exhibit a broad range of antimicrobial and antiviral activities. PG-1 form a well-defined structure in solution composed primarily of a two-stranded antiparallel beta sheet, with strands connected by a beta turn. Structure is stabilized by intrachain disulfide bonds. A nuclear magnetic resonance (NMR) study has determined the monomeric structure of PG-1 in solution [96 ]. In membrane environment PG-1 structure is membrane dependent: in bacteriamimetic anionic lipid membranes the peptide forms oligomeric transmembrane β-barrels, whereas in cholesterol-rich membranes mimicking eukaryotic cells the peptide forms β-sheet aggregates on the surface of the bilayer [97].

**Defensins** are small cationic proteins with molecules stabilized by several (usually three) disulfide bonds. In addition to antimicrobial activity they show other activities. These activities include chemotaxis, angiogenesis, modulation of adoptive immunological reactions, pro-inflammatory effects, cancer metastasis, etc. There are two main subfamilies of defensins: α and β. Both defensins consist of a tripled-stranded β sheet with a distinctive fold and so are natural extensions of the two-stranded β-hairpin AMPs such as protegrins. But compared to protegrins, defensins have weaker antimicrobial activities.

Human α-defensins are crystallized as dimmers. [98] The significance of dimerization for biological properties of a-defensins is as yet not completely understood. Human β-defensin in aqueous solution is monomer [99]. Both defensin monomers 3D structure shows amphiphilic character. Therefore, the mode of antibiotic action of defensins was thought to result from electrostatic interaction between the positively charged defensins and negatively charged microbial membranes, followed by unspecific membrane permeabilization or pore-formation. Using solid-state NMR spectroscopy, Zhang and coworker [100] have investigated the behavior of a human α-defensin ( HNP-1) in DMPC/DMPG bilayers. They show that membrane-bound HNP-1 exhibits a similar conformation to the water-soluble state. The protein is predominantly dimerized at high protein/lipid molar ratios and exhibits concentration-dependent oligomerization and membrane-bound topology. These data strongly support a "dimer pore" topology of α-defensin in which the polar top of the dimer lines an aqueous pore while the hydrophobic bottom faces the lipid chains. However, research during the past decade has demonstrated that defensin activities can be much more target-specific and that microbe-specific lipid receptors are involved in the killing activity of various defensins [101]

**Melittin** is a well-studied linear, cationic peptide from the venom of the European honey bee with antimicrobial and hemolytic activity. Although at low concentration, melittin is monomeric and adopts essentially a random coil conformation in aqueous solution; high salt, high melittin concentration and high pH promote an aggregation of monomeric melittin to a tetramer [102]. The effect of increasing melittin concentration on tetramer formation has been supported by an increase in the helical content of the peptide [103]. It has been shown that the binding of melittin to erythrocytes as a monomer is necessary for its hemolytic activity [102]. This is based on the observation that monomeric melittin is fully active whereas the tetrameric melittin, as induced by high phosphate counter ion concentration, lacks such activity under identical conditions. The aggregation state of melittin in membranes is important since this property is presumed to be associated with the function of melittin. Lipid composition and phase separation determines the aggregation behavior and pore formation of melittin. Melittin forms pores only in zwitterionic membranes and not in negatively charged membranes [104].

**LL-37 is** also linear, cationic, amphipathic, α-helical, antimicrobial peptide included into cathelicidin family. Its conformation in aqueous solution is sensitive to salt concentration, assuming a random coil configuration in pure water and becoming α-helical in the presence of millimolar anion concentrations [105]. The formation of helical structure also correlates with LL-37 aggregation and activity, which occurs at micromolar peptide concentrations in the presence of anions [105]. This helical, aggregated form of LL- 37 in solution is very different from other amphipathic, α-helical, antimicrobial peptides that are monomeric and unstructured in aqueous solution and only become α-helical upon association with a membrane.

It is generally assumed that the most linear antimicrobial peptides are unfolded in solution and get their active conformation only upon binding to their target bilayer [106]. Environmental conditions, as well as composition and physical state of the phospholipid bilayer may therefore be crucial for the peptide structure and aggregation. Although the most linear amphipathic antimicrobial peptides operate through random coil-to-helix structure transition upon interaction with microbial membrane [107], a growing number of studies on some CAPs have demonstrated that structural diversity of the peptides in the vicinity of membranes may lead to alternative mechanisms of action. For instance, despite very similar amino acid sequences, **Plasticins** adopted various structures at anionic and zwitterionic membrane interfaces including helices, destabilized helix states, β-hairpin, β-sheet and disordered states [108]. Cationic Plasticins are mainly helical when bound to anionic phospholipid vesicles, but the contribution of β-sheet structures increases when bound to zwitterionic vesicles. Plasticin, which has no net charge, adopts a helical structure when bound to anionic vesicles but is predominantly β-sheeted in the presence of zwitterionic phospholipids [108]. The sequence of plasticins encompasses three GXXXG motifs, known to mediate interactions between helical transmembrane domains of membrane proteins or helical fusion peptides [76]. But it has been shown, that GXXXG motifs is not sufficient to promote peptide oligomerization. Other factors may to be involved such as the sequence context and the intrinsic conformational flexibility. The same is possible for **bombinin H** and Alzheimer peptide because of the expected structural flexibility conferred by the glycines [81]. Structural variability and the potential to form higher order structures by the interaction with membrane thematically link bombinin, Plasticins and Alzheimer peptide Aβ and enable them to participate in cooperative processes. These higher order structures may either be helix assemblies or β-sheet oligomers [109].

**Tryptophan- and arginine-rich antimicrobial peptides** are membrane bound peptides that go far beyond regular alpha-helices and beta-sheet structures. [110]. **Indolicidin** was the first Trp-rich antimicrobial peptide discovered and has consequently been studied very thoroughly. It is a short, 13 amino acid peptide that is amidated at the C-terminus and contains the highest proportion of Trp [111]. Indolicidin belongs to the cathelicidin family of antimicrobial peptides. The structure of indolicidin was solved by NMR in neutral DPC and negatively charged SDS micelles [112]. The peptide did not adopt any classical secondary structure. Rather, the peptide displayed an extended conformation in both micelle types with β-turns being the most prominent structural motifs that were found [112]. The peptide forms a wedge-type shape, with hydrophobic Trp separated from positively charged regions of the peptide, that is, it adopts amphiphilic conformation [112]. It is unlikely that indolicidin acts in a barrel stave fashion, since it does not cause cell lysis at concentrations four times the MIC, indicating that indolicidin must possess another mode of action to kill bacteria [113].

**Tritrpticin** is another cathelicidin antimicrobial peptide. It is 13 amino acids long and has three consecutive Trp in its sequence [114]. The peptide was found to be structurally poorly defined in Tris buffer, but it adopted an stable, amphiphilic conformation in SDS micelles, resembling a wedge shape with Trp in the narrow part of the structure and with the hydrophilic side of the peptide which is made up of the Arg [115]. The amphipathic arrangement must therefore provide an energetic advantage, considering that the peptide prefers irregular conformation over regular secondary structure and corresponding hydrogen bonds. Tritrpticin induces membrane leakage to various degrees in different types of model membranes [116]. Fluorescence experiment indicates that tritrpticin inserts into negatively charged membranes more strongly and therefore has a greater lytic effect [116]. Such membrane insertion by tritrpticin may eventually lead to toroidal pore formation at high enough peptide concentrations [117].

**Proline-rich antimicrobial peptides** (PrAMPs) are a group of cationic host defense peptides of vertebrates and invertebrates characterized by a high content of prolines, often associated with arginines in repeated motifs. PrAMPs show a similar mechanism and selectively kill Gram-negative bacteria, with a low toxicity to animals. Unlike other types of antimicrobial peptides, their mode of action does not involve the lysis of bacterial membranes but only penetration into susceptible cells, where they then act on intracellular target [118].

**Drosocin, pyrrhocoricin, and apidaecin,** representing the short (18-20 amino acids) proline-rich antibacterial peptide family, originally isolated from insects and act on a target bacterial protein chaperone DnaK [119]. The proline-rich peptide family had multiple functions and functional domains, and perhaps carried separate modules for cell entry and bacterial killing. The Pro-Arg-Pro or similar motif repeats assist the entry into the host and subsequently into bacterial cells without any potential to destabilize the cells, and therefore without toxicity to eukaryotes. The antibacterial activity of the native products is provided by the independently functioning active site, capable of binding to the bacterial DnaK and preventing chaperone-assisted protein folding [120]. Proline-rich cell penetration modules may be general for antibacterial peptides in nature. For instance, the cathelicidin hydrophobic tail sequences reveal strong similarities to C terminal tails of pyrrhocoricin, drosocin or apidaecin [120].

Uversky et al. have shown that the combination of low mean hydrophobicity and relatively high net charge are an important prerequisite for the absence of regular structure in proteins under physiologic conditions, thus leading to "natively unfolded" proteins [121]. Then the view have been offered, that unfolded peptides and proteins have a strong tendency to poly-proline II (PPII) conformation locally, while conforming statistically to the overall dimensions of a statistical coil [122]. The PPII helix is often observed in the context of proline rich sequences, but sequences that are not enriched in proline can adopt this structure also. It has been shown, that peptides made up entirely of arginines or a TAT peptide, an arginine rich cell-penetrating peptide form PPII type conformations in membrane. An arginine rich peptides are able to interact and permeabilize biological membranes by forming the transient pores only, without cell lysis [123]

It can be suggested, that linier CAPs which are disordered in aqueous solution and thought to be in a statistical coil state may in fact be flickering in and out of a metastable PPII helical conformation. In membrane environment many of them form high ordered structures (alpha helical, beta structured or even aggregated) and so cause membrane disturbances by permeabilization in it. But some of them, for example proline rich peptides do not behave so, as they are not capable to form high order structures in membrane and remains in ppII conformation in latter environment. Consequently pro-rich peptide shows membrane penetrating capability only and finds targets for antimicrobial functioning inside a cell, with which forms stable complexes. In this respect, PrAMPs resemble Arg rich penetrating peptides.

It should be noted, that charged residue and Pro is the worst aggregator in both alpha and beta self-aggregation [124]. Pro-rich peptide is capable to aggregate only with conjugated to Gly form as it take place in the case of collagens or other Pro/Gly rich polypeptides [125].

**Gly rich antimicrobial peptides (GRAMP)** as a rule are multi-domain peptides with glycine rich motif as the particular among other motifs. In many cases function of the glycine-rich motifs is not well defined but in particular cases antimicrobial activity for separated glycine-rich part of the peptides has been shown. For instance, $SK_{66}His$ Glycine-rich peptide from Drosophilla CG13551 gene showed significant activity against Gram-positive bacteria [126]. As mentioned above, Gly play an intriguing role in peptide/protein structure where they can act as tightly packing amino acids with flexible main chain. It means that Gly is the best "aggregator" residue. Consequently we can speculate, that mechanisms of action of GRAMP are based on peptide aggregation. Indeed, glycine-rich peptide Pg-AMP1 from guava seeds acts by formation of a dimmer. Pg-AMP1 shows sequence similarity to plant glycine-rich protein family and a 3D structural homology to an enterotoxin from Escherichia coli, and other antibacterial proteins [127]

Thus we can conclude that generally, cationic antimicrobial peptides show structural variability in aqueous solution as well as in membrane environment. Majority of CAPs have propensities to flexibility, disordering and non-aggregation in aqueous solution, whereas membrane promotes the peptides ability to form high order structures. The type of the high order structure generated in membrane mainly depends on lipid composition and particular features of AMP sequence, for instance hydrophobicity, amphiphilicity, peculiar distribution along the chain small, hydrophobic and charged amino acids. The results of such peptides and membrane interactions may be the high membrane permeability, its disruption or formation of pores in it. Such kind of CAPs could be considered as "permeated" membrane peptides. But there are CAPs that don't form high order structures in membrane environment and have small effects on microbial membrane integrity. They often show flexible, unstable structures in both aqueous solution and membrane and so can easily translocate across microbial membranes. These "penetrating" through membrane CAPs as a rule have intracellular targets.

**Residue pairs responsible for CAP structure and aggregation**

As above mentioned, mechanisms of aggregation of transmembrane peptides in membrane environment and soluble peptides in polar environment differ but tendency of necessity of occurrence of particular pairs and triplets of amino acids (small, hydrophobic, beta branched, etc) in aggregating peptides is evident. For instance alternating patterns of polar and non-polar amino acids drive to beta aggregation in polar environmant, whereas pairs formed by two small or beta branched residues at register 4 affects on alpha aggregation in apolar environment.

In our laboratory analysis of the distribution of pairs formed by particular group of amino acids at any registar from 0 to 17 have been done [personal unpublished data]. Dataset of sequences of antimicrobial peptides for analysis was generated on the basis APD as linier CAP [50,128]. Hydrophobic beta-branched (HB), hydrophobic non beta-branched (HN), positively charged (PC) and small side chain (SG) amino acid groups were examined.

Statistics of pairs distribution was characterized by observed frequencies $f^k_i$, expected frequencies $F^k_i$ and standard deviation $D^k_i$. The observed frequencies were defind for measured dataset of CAP as $f^k_i = N^k_i / N$, where $N^k_i$ is number of sequential pairs of $k$ type amino acids ( k=HB, HN, PC, SG where HB= I, V, F; HN= L,M,W,C; PC= R,K; SG= G,A,S) with i number of other type residues between them and $N$ - sequential pairs of $k$ type with any number of residues between them. The expected frequencies $F^k_i$ and standard deviation $D^k_i$ were estimated via simulations. Simulation trials were performed by creating a random sets of sequences with the same lengths and composition as those in the measured dataset of CAP but with residues randomly shuffling. Standard deviations were computed based on 10 000 such trials.

Results of analysis are presented on Fig 1-4, where solid line corresponds to expected frequencies $F^k_i$ vs i dependence; dashed lines: $F^k_i +3 D^k_i$ vs i (upper) and $F^k_i -3 D^k_i$ vs i (lower) dependence; dots corresponds to observed frequencies. As we see both sequence and spatial motifs, that is pairs of HB and PC type at a register 4 are products of selection pressure on CAPs throughout evolution, either for structural integrity or for biological function. Results allow to consider possibilities of occurrence of alpha aggregation mechanisms for linier CAP in membrane environment and in this relation linier CAP resemblance to transmembrane peptides.

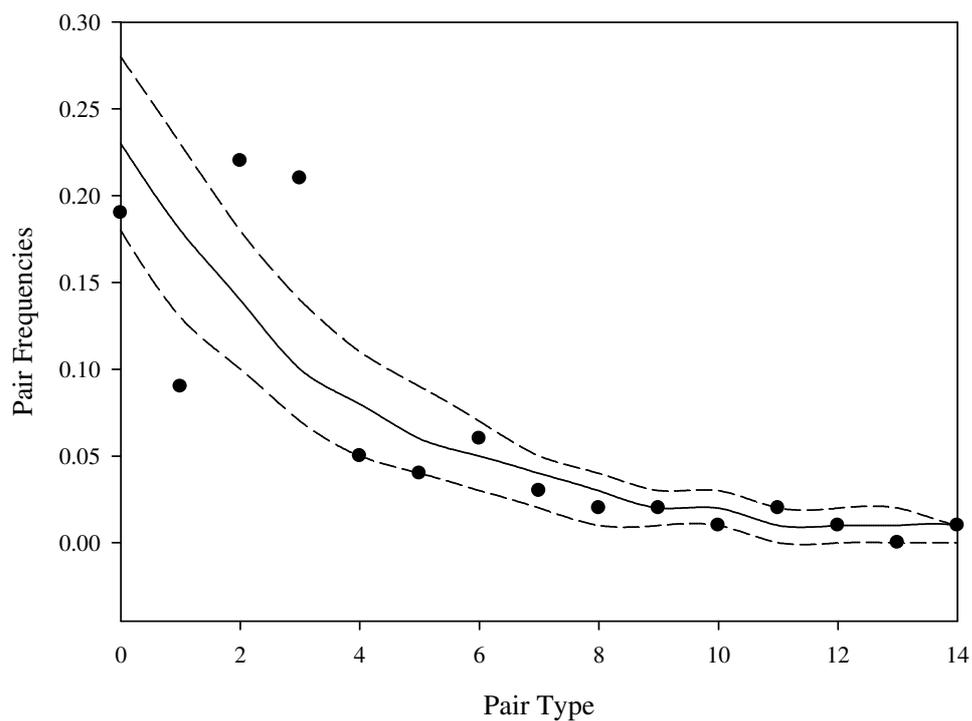

**Fig 1**. Statistics of pairs distribution for hydrophobic beta-branched amino acids. Solid line corresponds to expected frequencies $F^k_i$ *vs i* dependence; dashed lines: $F^k_i + 3 D^k_i$ vs i (upper) and $F^k_i - 3 D^k_i$ vs i (lower) dependence; dots corresponds to observed frequencies.

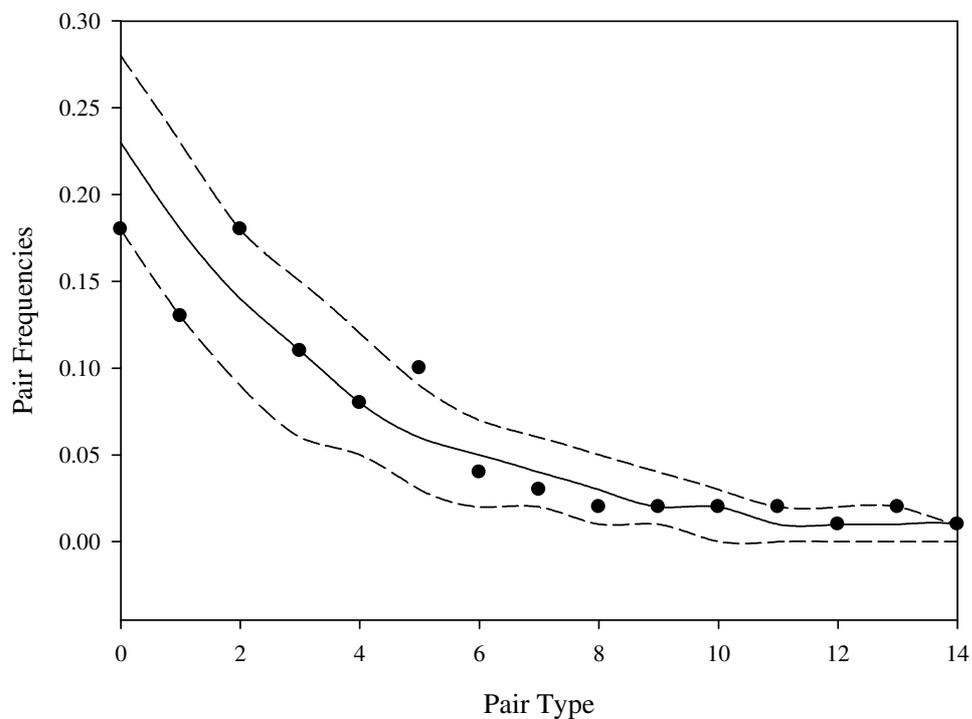

**Fig 2**. Statistics of pairs distribution for hydrophobic non-beta-branched amino acids. Solid line corresponds to expected frequencies $F^k_i$ *vs i* dependence; dashed lines: $F^k_i + 3 D^k_i$ vs i (upper) and $F^k_i - 3 D^k_i$ vs i (lower) dependence; dots corresponds to observed frequencies.

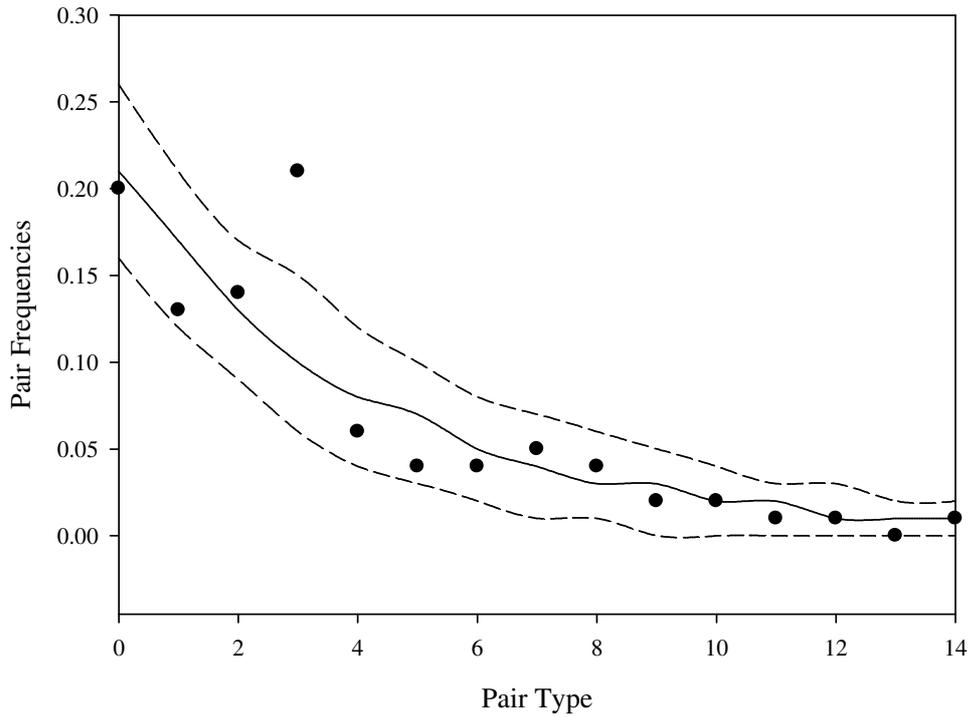

**Fig 3**. Statistics of pairs distribution for positively charged amino acids. Solid line corresponds to expected frequencies $F^k_i$ vs i dependence; dashed lines: $F^k_i +3 D^k_i$ vs i (upper) and $F^k_i -3 D^k_i$ vs i (lower) dependence; dots corresponds to observed frequencies.

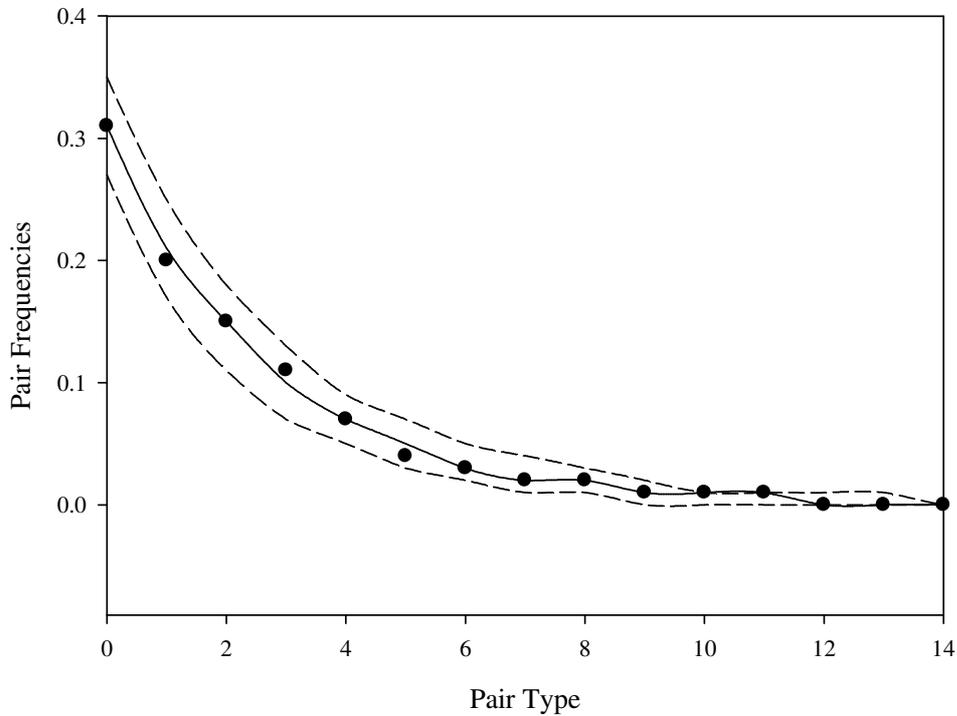

**Fig 4**. Statistics of pairs distribution small side chain amino acids. Solid line corresponds to expected frequencies $F^k_i$ vs i dependence; dashed lines: $F^k_i +3 D^k_i$ vs i (upper) and $F^k_i -3 D^k_i$ vs i (lower) dependence; dots corresponds to observed frequencies.

**Hydrophobicity and Amphiphilicity**

As a result of analysis of the characteristics of a novel family of synthetic cationic CAPs Glukhov et al. reveal two different threshold for peptide insertion in anionic and zwitterionic membranes [129]. The ''first threshold'' was defined as the minimal average segmental hydrophobicity required for peptide insertion from water into micelles or anionic bilayer membranes, whilst a second hydrophobicity threshold determines peptide's capability to insert into zwitterionic membrane. Stark et al. show that peptides become inactive as their average segmental hydrophobicities fall below the first threshold value, but sequences with increased cationic character (those rich with Lys, Arg,) convert the peptides into amphipathic molecules that likely have the capacity to embed effectively into the bacterial membrane surface [130]. When the average segmental hydrophobicities overcome second threshold peptides gain capability of insertion into eukaryotic membrane [130].

Lois M. Yin et al [131] examined four bioactive peptides, selected to investigate the extremes of hydrophobicity and positive charge distribution. They tried to answer the questions: "(1) why does increasing hydrophobicity lead to poorer antimicrobial activity and greater hemolytic toxicity; (2) whether altered charge distribution would improve the activity of CAPs with the same hydrophobicity level; and (3) which factor (hydrophobicity or charge distribution) is ultimately the more important contributor to effective CAP design and bioactivity." They have found that a CAP with relatively higher hydrophobicity undergoes a structural transition in the bacterial-type membranes from α- helix to β-type structure, whereas the corresponding CAP with lower hydrophobicity largely retains its α-helical conformation upon entering the membrane (129). This phenomenon is likely attributable to a charge neutralization effect as the peptides bind to the surface of the anionic bacterial membranes, and the resulting dehydrated environment facilitates of the formation of β-type aggregates of the CAP with a sequence of higher hydrophobicity . This result indicates that high segmental hydrophobicity can lead to an increased potential of peptide self association at the membrane surface and possibly of precipitation - thus limiting the concentration of peptide actually impacting on the bacterial membrane, and consequently reducing antimicrobial activity [131]; the latter notion is supported by the relatively lower MIC value for the peptide of low hydrophobicity vs. the peptide of high hydrophobicity.
On the other hand, Lois M. Yin et al 's results further indicate that when positive charge is distributed equally at both termini of the peptide aggregation above dimers is eliminated. The separated charge distribution of CAPs – which otherwise retain identical hydrophobic cores with AXXXA motif(s) - does not disrupt their dimerization ability in hydrophobic environments, and accordingly, that primary sequence motifs remain the chief determinant of oligomeric status [131].
In the end authors concluded that maximal antimicrobial activity and minimal hemolytic activity of CAPs do not involve only one factor, but rather require a good balance among (i) peptide helicity, (ii) optimal hydrophobicity of the core segment; (iii) positive charges and their distribution; (iv) dimerization and/or oligomerization ability in the membrane; and (v) minimization of aggregation in aqueous solution [131].

In order to determine what characteristics distinguish efficiently linier CAP (LCAP) from other peptides, (other membrane-interactive or non-functional (random)) Vishnepolski and Pirtskhalava [50] have performed comparative analysis of sequences of the three sets of peptides: LCAP, transmembrane (TMP) and randomly selected fragments from the soluble proteins (RFP). Results of analysis show, that a) the LCAP has lower average hydrophobicity than the transmembrane helices, but greater than random fragments from the soluble proteins; b) Linear hydrophobic moment ( "linier amphyphilicity") values distribution suggests that for the most antimicrobial peptides there is no significant linear separation of hydrophobic and hydrophilic residues along the peptide chain. At the same time in the transmembrane peptides linear separation of hydrophobic and hydrophilic group of residues occurs. c) LCAP penetrate into the membrane at a shallow depth (8-15Å) parallel to the membrane surface. On the other hand, for the trans-membrane fragments it is energetically more favorable to penetrate more deeply into the membrane in perpendicular to the membrane surface orientation. (d= 2-10 Å). Peptides from the dataset of random protein fragments are located closer to the membrane surface; d) Eisenberg's hydrophobic moment (" spacial amphyphilicity") [48 ] is the best separator between non-antimicrobial and antimicrobial peptides as the LCAP possess the highest spacial amphyphaticity.

## Conclusions on commonness (generalities) and distinctions

Membrane-penetrating CAPs as well as a cell-penetrating peptides (CPP) generally can be divided into two categories: rich in proline or positively charged amino acids and amphipathic (for example indolicidin). On the other hand, hydrophobicity and amphiphaticity are common features of the "permeated" the membrane CAPs.

Generally CPP and CAP are very similar peptides. Short cationic structure and the ability to interact with membrane is of crucial importance to both. So, depending on peptide lipid ratio, CAP can behave as CPP and vise versa. For instance, at high enough concentration, peptides reported as CPP perturb membranes and become membrane permeabilizers [132], whereas CAPs may reach cytoplasmatic targets before membrane permeabilization, (i.e. at a low concentration) [133].

Thus, CAP, CPP peptides belong to the same class of membrane active peptides and their differentiation is difficult or even may not be right. At the same time the revealing of the common features of the CAP sequences that gives us possibilities to distinguish CAPs from transmembrane and average soluble peptides is the real task and its solution is valuable for antimicrobial peptide prediction and design.

Rich in proline and positively charged amino acids CAPs are easily distinguishable from transmembrane peptide or average soluble peptide by amino acid composition only. The differentiation of amphiphatic CAPs from transmembrane (TMP) or average soluble peptides (ESP) is more difficult as it seems, because CAPs posses some features making peptides prone to aggregation as it do TMP and ESP. For instance some level of hidrophobicity and occurrence of certain amino acid motifs (for instance "small amino acids zippers" in bombinin, Plasticins,Alzheimer peptide, etc) pull CAPs to aggregation in membrane environment, at the same time aphiphaticity is a main driving force for aggregation in water environment. But there are factors that resist CAP aggregation. Such factors are abundance of bulky (I,V,W) and charged (R,K) amino acids that protect from aggregation in water. The charged amino acids weakens aggregation stabilizing interactions in membrane also [6]. In the certain membrane environment just the balance between hidrophobicity and charge is the main determinant which gives transmembrane peptides capability to form a stable aggregates. At the same time the balance between the same physicochemical characteristics is a cause of a resistance of some CAPs to stable self-aggregation or just aggregation [6]. Peptide amino acid composition (hydrophobicity) and distribution along the chain ("linier amphiphaticity") or on the surface of 3D structure ("spacial amphyphaticity") of non-polar and charged amino acids determine peculiarities of interactions with lipid bilayer. For instance high hidrophobicity and linier amphyphaticity predispose transmembrane peptides to interact with membrane along their whole width and to form stable complexes with lipid molecules and with each other in perpendicular to membrane surface orientation. The CAPs, on the other hand, because of its high "spacial amphyphaticity" and non-high hydriphobicity preferably situate in the interfacial area of membrane, at parallel to membrane orientation. Positive charge gives CAPs selectivity to procariotic membrane. Interaction with membrane in parallel orientation at the initial stage of action is a common feature of CAPs [6,128]. But the farther development of events depends on sequence of particular peptides, their concentration and the composition of membrane. Generally, events can go in different scenario. For instance, there are peptides which act by a) "barrel stave" mechanism, when self-aggregates are formed and during the process of the aggregation peptides change their parallel orientation to the perpendicular one [134 ]; b) "torroidal pore" mechanism when high-ordered structure which consists of peptide and lipid molecules are formed. Peptides orientation are changed also;[135]  c) by "carpet" mechanism  which means formation of peptide lipid aggregates only [136]; d) penetrating  membrane mechanism when membrane damage does not take place and targets are situated in cytoplasm [120,137], etc . In any case common features of CAPs: capability to interact with lipid belyer and such factors as sequence of particular peptide, their concentration and concrete composition of membrane determine capability of high order structure formation and membrane permeabilization.

Taking into account the abovementioned we can reveal the following features of CAPs sequences:

a) **The amphiphatic CAPs** are characterized by moderate hydrophobicity, high hydrophobic moment, low propensity to aggregation in aqueous solution and more or less propensity to aggregation in membrane (aggregates are unstable). Sequence defines parallel to membrane surface orientation.
b) **The transmembrane peptides are** characterized by high hydrophobicity, low hydrophobic moment and high propensity to aggregation in membrane (aggregaties are stable). Sequence defines perpendicular to membrane surface orientation.
c) **The average soluble peptides** are characterized by low hydrophobicity, high hydrophobic moment, and high propensity to aggregation in aqueous solution.
d) An aggregating mechanisms in membrane environment are alike for membrane and antimicrobial peptides, although an occurrence of such factors as abundance of charged amino acids or proline diminish stability of CAP's aggregates or even causes failure in CAP's aggregation.

Consequently we can conclude that

a) antimicrobial peptide resembles transmembrane one in aggregation mechanisms but highly differ in: hydrophobicity and amphiphaticity, stability of aggregates in membrane and orientation.

b) average soluble peptide highly resembles antimicrobial one in amphiphilicity and less in hydrophobicity, but differ in propensity to aggregation in aqueous solution.

It is clear that there are exclusions from latter generalities. But the existence of such common characteristics for each class of peptides (transmembrane, amphiphatic CAP and "average soluble") and consequently certain likeness and differences between them can be helpful in the experiments of design of new antimicrobials or in development of efficient, computer-aided, antimicrobial peptide prediction method.

# Acknowledgement

The designated work has fulfilled by financial support of the CNRS / SRNSF Grant No 09/10.